# Overview of Spintronic Sensors, Internet of Things, and Smart Living

X. Liu[1], K. H. Lam[1,*], K. Zhu[1], C. Zheng[1], X. Li[1], Y. Du[1], Chunhua Liu[2], P. W. T. Pong[1,*]

[1]Department of Electrical and Electronic Engineering, The University of Hong Kong, Pokfulam Road, Hong Kong

[2]School of Energy and Environment, City University of Hong Kong, Tat Chee Avenue, Hong Kong

*Correspondence: khlam@eee.hku.hk; Tel.: +852 2857 8401 (K. H. Lam)

*Correspondence: ppong@eee.hku.hk; Tel.: +852 2857 8491 (P. W. T. Pong)

**Abstract**

Smart living is a trending lifestyle that envisions lower energy consumption, sound public services, and better quality of life for human being. The Internet of Things (IoT) is a compelling platform connecting various sensors around us to the Internet, providing great opportunities for the realization of smart living. Spintronic sensors with superb measuring ability and multiple unique advantages can be an important piece of cornerstone for IoT. In this review, we discuss successful applications of spintronic sensors in electrical current sensing, transmission and distribution lines monitoring, vehicle detection, and biodetection. Traditional monitoring systems with limited sensors and wired communication can merely collect fragmented data in the application domains. In this paper, the wireless spintronic sensor networks (WSSNs) will be proposed and illustrated to provide pervasive monitoring systems, which facilitate the intelligent surveillance and management over building, power grid, transport, and healthcare. The database of collected information will be of great use to the policy making in public services and city planning. This work provides insights for realizing smart living through the integration of IoT with spintronic sensor technology.

## I. Introduction

The rapidly growing population, increasing demands on quality of life, and rising pressure on environmental conservation have been initiating the social reform for a more intelligent lifestyle. Smart living has been introduced as a trending style of living that will bring about more efficient energy consumption, better healthcare, and elevated standard of living through integrating advanced information and communication technologies (ICT), smart sensing technology, ubiquitous computing, big data analytics and intelligent decision-making. Smart living will be primarily accomplished by the concept of smart city [1]. The final goal of the smart city is to enhance the utilization efficiency of various public resources, promoting the quality of services offered to the citizens while reducing the operational cost of public infrastructure and administrations [2]. Smart building, smart grid, smart transport, and smart healthcare are the key application domains to support smart city vision [3]. Researches focusing on these areas have obtained considerable progress in the past decade; nonetheless, there still exist roadblocking challenges in these regimes due to the novelty and complexity of smart living, such as improving energy efficiency of buildings without reducing the comfort level [4], integrating renewables and self-healing for power systems [5], highly-efficient management of traffic congestion [6], and provision of quality healthcare for growing populations at reduced overall costs [7]. Such challenges need to be tackled by new monitoring systems rather than the traditional ones that function with limited sensors and wired communications.

The Internet of Things (IoT) is a compelling paradigm to support smart living by means of connecting a great quantity of digitally augmented physical objects to the Internet. These objects, especially sensors are being connected everywhere and at all time collecting the information of interest. The wireless sensor network (WSN) based IoT platform is popular in modern monitoring systems. It is predicted that the number of sensors in diverse applications will exceed one trillion by 2022 and over 100 trillion by 2030 [8]. Over the last few decades, spintronic sensors working as solid-state magnetic sensors have gained continuous attention and research effort because of their traits such as superb sensitivity, low power consumption, compactness, wide bandwidth, room-temperature operation and CMOS compatibility. In this paper, the wireless spintronic sensor network (WSSN) is proposed to seamlessly integrate the spintronic sensors into the IoT. This winning combination can be an important piece of cornerstone for IoT to realize smart living in the 21st century.

While the previous reviews on the spintronic sensors focus on the development of their structures and industrial applications [9, 10], there still lacks an overview and outlook on the potential roles that spintronic sensors may play in the realization of smart living. The objective of this paper is to bridge the gap between the scientific research on spintronic sensors and the realization of the IoT-based smart living. This paper is organized as follows: Section 2 presents an overview of the principle of spintronic sensors and their sensing capabilities relevant to electrical current sensing, transmission and distribution lines monitoring, vehicle detection, and biodetection. In Section 3, we briefly introduce the IoT concept and architecture. Specifically, we explain how to organize a WSSN to support the IoT. In Section 4, the WSSN based or related applications to support smart building, smart grid, smart transport, and smart healthcare are discussed, respectively. Finally, Section 5 provides the future outlook and concluding remarks.

## II. Spintronic sensors

To achieve smart living, abundant types of data are required to be detected, collected and connected to the Internet. Spintronic sensors can provide measurement data of magnetic field from which magnetic-field-related parameters of some real-world objects can be derived. The database of "Things" can then be enriched, facilitating the development of the IoT-enabled smart living. In this section, we will introduce the basic principles of spintronic sensors and discuss their sensing capabilities relevant to smart living.

### 2.1 Basic principles of spintronic sensors

Typically, the resistance of spintronic sensors depends on both the magnitude and direction of the external magnetic field, known as the magnetoresistance (MR) effect. Based on distinct underlying mechanisms, spintronic sensors are normally categorized into anisotropic magnetoresistance (AMR), giant magnetoresistance (GMR), and tunneling magnetoresistance (TMR) sensors. The general schemes of these spintronic sensors are illustrated in Fig. 1.

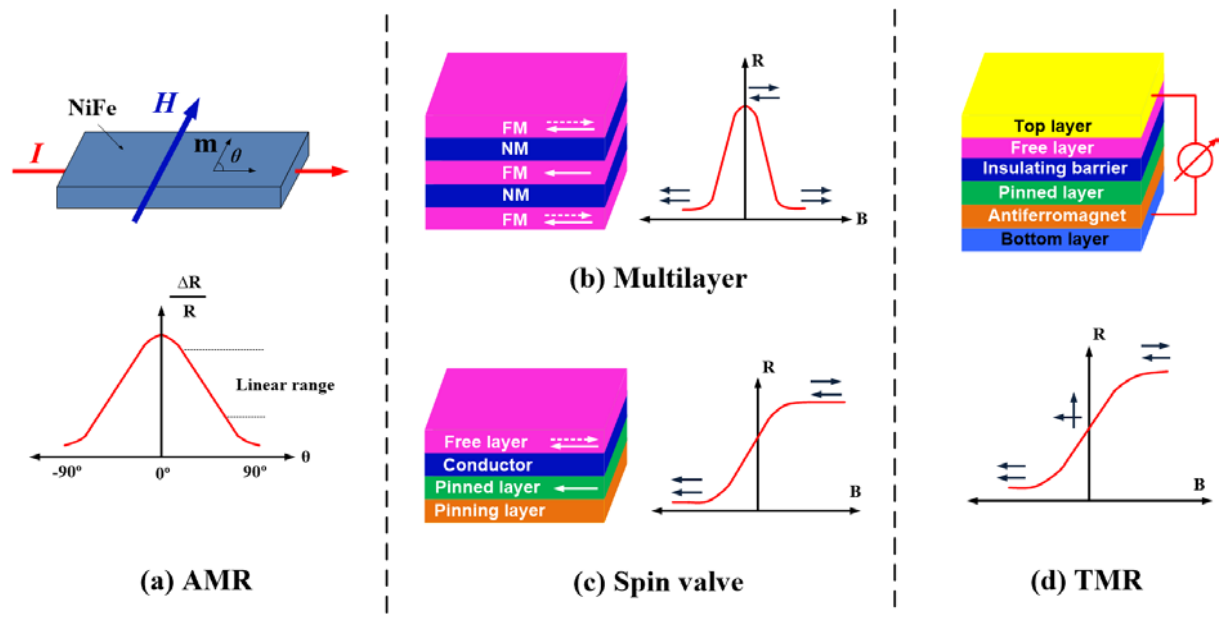

**Fig. 1** Basic structures of AMR, GMR, and TMR sensors

### 2.1.1 Anisotropic magnetoresistance (AMR)

The AMR effect in ferromagnetic metals and alloys stems from the anisotropic scattering of conduction electrons induced by spin-orbit interaction. The spin-orbit interaction leads to a mixing of spin-up and spin-down electron states in the ferromagnetic materials. Such a mixing of electron states is affected by the magnetization direction of the material. This behavior gives rise to a magnetization-direction-dependent s-d scattering rate, which influences the resistance of the AMR system, as shown in Fig. 1(a). Since the magnetization direction of the AMR material can be changed by the magnitude and direction of an external magnetic field, a link between the resistance of AMR sensors and the external magnetic field can be established. Based on this principle, the information of magnetic field (e.g., direction, magnitude) can be obtained by using AMR sensors. Furthermore, the directional sensing ability of the AMR sensor can be utilized to determine the rotational conditions (e.g., angle, rotational speed) of a magnetic object. As a result, nowadays AMR sensors can work as contactless magnetometers, angle or rotation sensors.

### 2.1.2 Giant magnetoresistance (GMR)

The giant magnetoresistance effect in ferromagnetic/nonmagnetic (FM/NM) thin film stacks was discovered in 1988, for which Fert [11] and Grünberg [12] were awarded the 2007 Nobel Prize. Unlike the AMR sensor where the electron spins are randomly oriented, the FM layer in GMR sensors serves as a spin filter to polarize the spin of electrons. Hence, stronger spin-dependent scattering is expected when the spin current flows through the NM spacer to the interface of another FM layer. The resistance of GMR sensors is dependent on the relative orientation of the magnetizations of the neighboring FM layers. The GMR ratio is defined by:

$$GMR = \frac{R_{AP} - R_P}{R_P}, \tag{1}$$

where $R_{AP}$ and $R_P$ represent the resistance at antiparallel and parallel state, respectively. High GMR ratio is preferred for sensors with high sensitivity. Besides, the linear response of the GMR sensor defines the detectable field range while the coercivity determines the extent of hysteresis effect in the magnetization transfer curve.

Magnetic multilayers and spin valves are used in commercial GMR sensors. The magnetic multilayers are composed of repetitions of FM/NM layer pairs, as shown in Fig. 1(b). Under zero magnetic field, the multilayers exhibit high resistance since the FM layers are oriented in opposite direction through interlayer exchange coupling. When the magnetizations of the FM layers are gradually aligned to the parallel configuration by an external magnetic field, the resistance of this MR sensor decreases from the maximum resistance to the minimum value. A large magnetic field is required to overcome the strong exchange coupling amongst the FM layers, resulting in saturation magnetic field up to hundreds of mT [9, 13]. However, the extended linear detectable field range is accompanied with a drawback of relatively low sensitivity. The development of spin valves [14, 15] can provide higher sensitivity through introducing a soft FM layer in the FM/NM/FM trilayer structures (see Fig. 1(c)). The soft FM layer can be freely rotated by the external magnetic field as the interlayer exchange coupling is reduced by the relatively thicker NM spacer. The hard FM layer is pinned by an antiferromagnetic (AF) layer through direct exchange coupling to maintain its magnetization direction when the reverse magnetic field is applied.

The recent efforts on the development of spintronic sensors have been devoted to increasing the GMR ratio and reducing the coercivity. Through engaging FM materials with high spin polarization ratio (such as Heusler alloy [16-18]), GMR ratio of 74.8% was achieved [19], which is much higher than that in conventional spin valves (5 ~ 10 %) [13]. The insertion of nano-oxide-layer as the NM spacer led to a remarkable increase in GMR ratio due to the enhanced spin-dependent scattering at the metal-oxide interfaces [20, 21]. On the other hand, the coercivity can be reduced through engaging soft FM materials (such as supermalloy or Conetic alloy [22, 23]) or synthetic ferrimagnetic structure [24] as free layers.

### 2.1.3 Tunneling magnetoresistance (TMR)

Compared to spin valves, a typical TMR sensor [25-29] has a similar basic structure but with the metallic spacer replaced by a thin (0.5 ~ 2 nm) insulating barrier, as shown in Fig. 1(d). As a result, instead of the spin-dependent scattering effect, the spin-dependent tunneling effect [30] is involved in TMR sensors, which are therefore called magnetic tunnel junctions (MTJs) [31]. The spin-dependent tunneling effect in MTJs can either be incoherent [32, 33] or coherent [25, 27, 34, 35], depending on the type of the tunnel barrier. The incoherent tunneling is typically exhibited in the AlO$_x$-based MTJ [36-39], whose TMR ratio is phenomenologically described by Julliere's model [32, 37, 40]:

$$TMR = \frac{2P_1 P_2}{1 - P_1 P_2}, \tag{2}$$

where $P_1$ and $P_2$ are the spin polarizations of two FM layers, respectively. In the AlO$_x$-based MTJ, the highest observed TMR ratio at room temperature is 70.4% [40], which corresponds to a spin polarization of 0.51 for the CoFeB FM layer based on Julliere's model. Considering the thermally-driven spin fluctuation at room temperature, such a spin polarization is close to the spin polarization (usually 0 ~ 0.6) of 3$d$ Co or Fe-based FM alloys measured in FM/AlO$_x$/superconductor junctions at an ultralow temperature (< 4.2 K). One of the breakthroughs with respect to the TMR effect is the use of oriented single-crystal MgO barrier, achieving the spin-dependent coherent tunnel in MTJs. In such MgO-based MTJs, the totally symmetric $\Delta_1$ Bloch states (spd hybridized states) in the FM layer can effectively couple with the evanescent $\Delta_1$ states in the MgO barrier and decay much slower than the $\Delta_2$ states (d states) and $\Delta_5$ states (pd hybridized states) [31, 41]. Therefore, only highly polarized $\Delta_1$ states travel across the MgO barrier, which leads to an extremely high spin polarization of the tunneling current and thus a large TMR ratio. By optimizing the fabrication process, TMR ratio of the MgO-based MTJ can even reach 604% at room temperature [35]. In addition, other new magnetic materials are developing to enhance the performance of TMR sensors, such as Heusler alloys [40, 42-45], ferrites [46, 47], rutiles [48], perovskites [49], dilute magnetic semiconductors [50], and so on.

For sensor applications, not only the sensitivity is required to be improved but also the noise level should be suppressed in order to boost up the signal-to-noise ratio (SNR). The MR ratio of TMR sensors is 1 ~ 2 orders of magnitude higher than that of GMR sensors [31], but the noise level of TMR sensors is much higher than that of GMR sensors [51]. Therefore, the main roadblock to achieving ultrahigh SNR of TMR sensors is their relatively high noise level as compared to GMR sensors. Intrinsic noises in TMR sensors include thermal noise [51], shot noise [52, 53], electronic 1/f noise [54, 55], magnetic 1/f noise [56, 57], and random telegraph noise (RTN) [58, 59]. In the sensing region of a TMR sensor, the dominating noise source is normally the magnetic 1/f noise, which is essentially attributed to the thermally-driven magnetization fluctuation of the free FM layer. Such magnetization fluctuation can be experimentally suppressed by applying a hard-axis magnetic field, annealing in a high magnetic field, or utilizing the voltage-induced anisotropy modulation, which therefore reduces the magnetic 1/f noise of TMR sensors.

### 2.1.4 Spintronic sensor fabrication

To yield a linear response, the easy axes of free and pinned layers of the spin valves and MTJs should be set orthogonal. The pinned layer is defined in a transverse magnetization direction through annealing in a strong magnetic field while the free layer keeps in the longitudinal direction [60]. To increase the detectivity, the MR elements are configured in a Wheatstone bridge (e.g., GMR sensor GF708 [61]) to provide differential output and integrated with magnetic flux concentrator (MFC) to enhance the signal. The state-of-the-art spintronic sensors exhibit the ability to detect ultimate weak magnetic fields (from nT down to pT) at room temperature [62-66], which fulfill the requirements of high sensitivity for smart-sensing applications.

## 2.2 Sensing capabilities of spintronic sensors relevant to smart living

Researchers have developed extensive sensing capabilities by the virtues of spintronic sensors. The following subsections will overview four typical capabilities of spintronic sensors, i.e., electrical current sensing, transmission and distribution lines monitoring, vehicle detection, and biodetection. They will contribute to the collection of several key parameters in Section 4 (e.g., power consumption/generation of electric applications, electrical and position parameters of transmission lines, traffic information, and health conditions).

### 2.2.1 Electrical current sensing

Magnetic field is emanated from a current-carrying conductor (see Fig. 2) and the magnitude of the magnetic field at a sensing point ($|\vec{B}|$) can be deduced from the Biot-Savart law as following:

$$|\vec{B}| = \frac{\mu_0}{4\pi} \int \frac{Id\vec{L} \times \hat{r}}{|\vec{r}|^2}, \tag{3}$$

where $d\vec{L}$ is the infinitesimal length of the conductor carrying current $I$, $\hat{r}$ is the unit vector to specify the direction of the vector $\vec{r}$ from the current to the field point, and $\mu_0$ is the magnetic permeability constant. The magnetic field ($\vec{B}$) can be decomposed into three mutually orthogonal axes as

$$|\vec{B_x}| = |\vec{B}|\cos\alpha, \ |\vec{B_y}| = |\vec{B}|\cos\beta, \ |\vec{B_z}| = |\vec{B}|\cos\gamma, \tag{4}$$

where $\alpha$, $\beta$, $\gamma$ are the respective angles formed between $\vec{B}$ and the corresponding sensitive axis of the sensor (i.e. X-axis, Y-axis, Z-axis). With the decoupled magnetic field measured, the resultant magnetic field can be obtained by

$$\vec{B} = |\vec{B_x}|\vec{a_x} + |\vec{B_y}|\vec{a_y} + |\vec{B_z}|\vec{a_z}, \tag{5}$$

where $\vec{a_x}$, $\vec{a_y}$, $\vec{a_z}$ are the unit vectors of the three mutually orthogonal sensitive axes of the sensor. The current of the conductor can therefore be derived from the relationship established in Eq. (3).

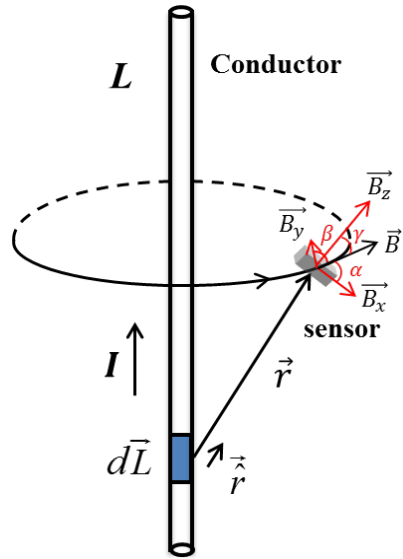

**Fig. 2** Magnetic field emanated from a current-carrying conductor.

Based on the above-mentioned principle, spintronic sensors are increasingly gaining popularity in current sensing compared to current transformers (CT), shunt resistors, Rogoswski coils, fluxgates and Hall effect sensors [9, 67, 68]. Though CTs have been applied in power systems for a long time, they can only measure alternating currents with limited frequency bandwidth and they are bulky in size. Shunt resistors interrupt the original circuit, and the intrinsic inductance limits its measuring sensitivity and bandwidth. Rogoswski coils have poor accuracy when sensing small and low-frequency currents, which restricts their broader application. Although fluxgate sensors can significantly improve the measuring accuracy, they are now mainly employed in the calibration systems, diagnosis, and laboratory equipment because of the large size and high cost. The low sensitivity of Hall effect sensors leads to the necessity of using a magnetic core to clamp around the current conductor for signal concentration which presents difficulty to practical deployment, particularly when the conductor is not isolated and thus cannot be clamped. In contrast, spintronic sensors show great potentials in electrical current sensing due to their high sensitivity, compact size, low price, AC and DC detection ability, broad frequency bandwidth, non-contact installation, and low power consumption.

The compactness and non-contact detecting make spintronic current sensors idea for the integrated circuit (IC) design. Nowadays current sensors are ubiquitous and essential in the IC design. For example, current sensors are placed on every power pin to ensure the homogeneous power distribution [69], or at the power connections of different blocks to indicate the levels of activity performed by the blocks for rescheduling the workload [70]. Due to the compact size of spintronic sensors, current sensing modules in ICs can be further miniaturized. Moreover, spintronic sensors do not interrupt the original circuit path and thus they do not degrade the original circuit performance or bring in power losses since it is non-contact measurement.

The DC detecting ability enables the spintronic sensors to measure DC current while the broad frequency bandwidth for AC detection is conducive to transient current monitoring. DC residential appliances such as televisions, computers, and LED light fixtures are very common in the 21st century.

It is envisioned to integrate DC buses into building electrical systems for connecting and accommodating the DC loads. Renewable resources (e.g., solar energy) and energy storage elements (e.g., batteries) are also inherently DC systems. Therefore, the DC detectability of the sensor is being critical. Various kinds of transient AC currents (e.g., lightning currents, switch impulse currents, and leakage currents) occur in a building, and they usually have large amplitudes and wide frequency bandwidth (e.g., lightning current can be as high as several kA and several MHz frequency). The broad frequency response of spintronic sensors makes them suitable for the tasks of measuring such AC currents.

Low cost and low power consumption of spintronic sensors are favorable for large-scale application and deployment. Nowadays the current sensors are almost installed in a centralized hub. For example, CTs are installed in the substations, and the electricity meters are installed at customers' premises to measure the total energy consumption for billing. With the advent of home automation, many residential appliances will be networked together and operated through the energy management system (EMS). The control of the electric appliances will not only be through on-and-off switches but also respond to the signals indicating the electricity usage condition of each device under EMS management. All these depend on a current sensing network on the appliances. Spintronic sensors have potential to be embedded in a large amount of the appliances at low cost with little power consumption.

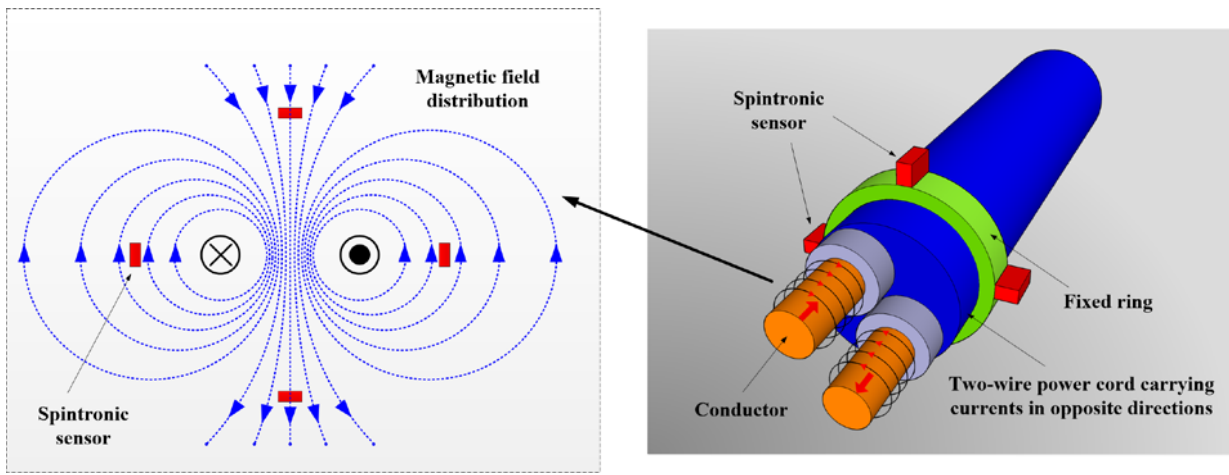

**Fig. 3** Spintronic sensor deployment for current sensing in two-conductor power cable [71]

Furthermore, spintronic sensors can measure the currents in multiple-conductor cables without need to remove insulation from the cable whereas traditional current clamps cannot measure the currents in multiple-conductor cables. In [71], four spintronic sensors encircle two-conductor cable of any type, and these four sensors are 90° rotationally symmetric around the central axis of the power cable (as shown in Fig. 3). Due to the symmetric magnetic field distribution for the currents in opposite directions, the magnetic field can be decomposed, and then measured by four spintronic sensors. Hence, the currents in power cable and the rotation angle of the power cable around its central axis can be resolved from the magnetic field measured. With this technique, spintronic sensors can be implemented to measure the current of power cables connected to domestic appliances or a household which are typically two-condutor cable. Meanwhile, non-contact voltage sensors are utilized to detect the real-time voltages of these cables, and thus the instantaneous power can be determined by multiplying current with voltage. Using this convenient technique, the power consumption of each individual electric appliance in a household or building can be monitored.

### 2.2.2 Transmission and distribution lines monitoring

Transmission and distribution lines are the critical elements in power system to transfer the electricity from the generation side to the consumption load over long distances. Overhead transmission lines are usually deployed for the transmission network while underground power cables are frequently used in the distribution network. Conventionally, the operating currents of the overhead transmission lines and underground cables are monitored by the CTs at the substations. Apart from the drawbacks such as limited measurement range and frequency bandwidth, CTs are costly (typically over US$ 100k), bulky (with volume size in the order of $m^3$), and need regular maintenance (e.g., insulation oil replacement). Great care must be taken to ensure the secondary side never open-circuited. Spintronic sensors overcome the above weaknesses of CTs applied in transmission and distribution lines monitoring systems.

In the previous researches, a non-invasive platform to monitor the current of overhead transmission lines and underground cables by employing spintronic sensors has been developed accordingly [72-74]. The configuration of the overhead transmission lines and the deployment of spintronic sensors are shown in Fig. 4(a). The three-phase high-voltage conductors of the overhead transmission lines are current-energized. The spintronic sensor arrays can be deployed either on the ground level [72] or installed on the transmission towers [75] (as sometimes the geographical condition does not allow them to be deployed on the ground). The magnitude of magnetic field at the ground level under a transmission line is typically on the order of $10^{-5}$ T while it is on the order of $10^{-4}$ T on the transmission towers, and currently spintronic sensors are sensitive enough to measure these transmission-line magnetic fields. The configuration of the underground power cables and the deployment of spintronic sensors are shown in Fig. 4(b). The spintronic sensor arrays are installed around the cable surface to measure the magnetic fields generated by the three-phase conductors of the cable. An array of spintronic sensors in a circle is installed around the cable surface to measure the comprehensive magnetic information. The magnitude of the magnetic field around underground power cable surface is at the level of mT, which can be detected by spintronic sensors. The currents can be reconstructed from the magnetic field measured by the stochastic optimization algorithm (refer to Fig. 7 in [72]). The stochastic optimization algorithm is adopted since the current cannot be solved easily from the magnetic field information in an analytical way. The reconstructed results resemble the actual ones with very small errors, as demonstrated in [72, 74].

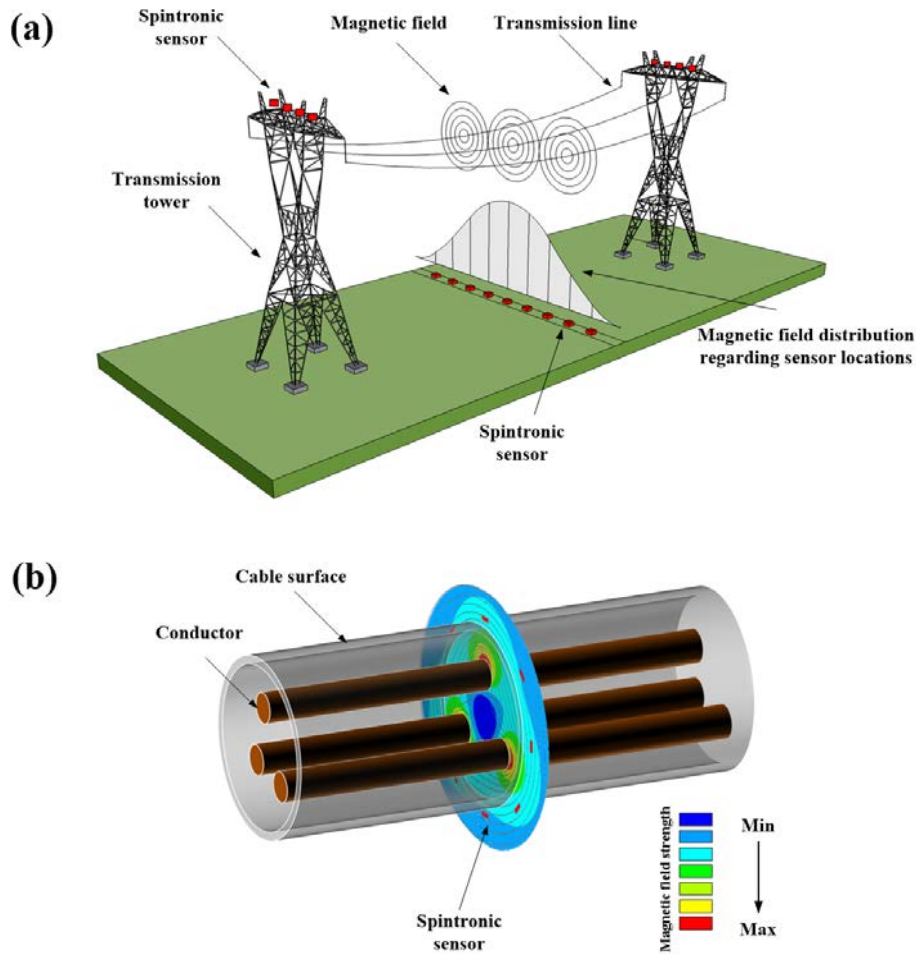

**Fig. 4** Spintronic sensor deployment in monitoring transmission and distribution lines. (a) The configuration of three-phase overhead transmission lines and spintronic sensors. (b) The configuration of underground power cable and spintronic sensors.

This platform is compact and low-cost by deploying spintronic sensors. Spintronic sensors do not need regular maintenance. The sensing platform is entirely isolated from the primary circuit, thus the reconstruction of currents and spatial parameters based on magnetic field sensing does not disturb the operation of transmission lines and underground cables. The spintronic sensors also show the enhanced measurement range and frequency bandwidth than traditional CTs [76]. By virtue of low cost and compact size, the platform can be extensively deployed over transmission lines to establish the sensing network [77]. It should also be noted that the developed platform excels the existing current clamps because it can carry out current sensing for each individual conductor in a multi-core cable whereas the current clamp measures the vector sum of all currents together and thus there would be a net current of zero if it clamps the three-phase balanced currents together. Moreover, the insulating layers of the multi-core underground power cable must be removed to make the conductors exposed if the current clamps are going to clamp around each conductor individually. The easy installation of the developed platform makes itself suitable for wide area deployment on underground power cables.

### 2.2.3 Vehicle detection

Intelligent transport system (ITS) integrates the information and communication technologies into the transport infrastructure, aiming to deliver innovative services for various modes of transport systems with improved safety, security, efficiency, mobility and environmental performance. In recent years, ITS has attracted widespread attention from both government and motor corporations. Vehicle detection, as the most fundamental element of the ITS, collects traffic information such as vehicle speed, occupancy rate, and traffic volume. Vehicle detectors based on spintronic sensors have been widely used for vehicle detection applications [78-80].

The Earth provides a uniform and stable magnetic field over the planet surface. A ferrous or metal object, like a vehicle, can be considered as a model consisting of a number of bipolar magnets with N-S polarization direction. A vehicle can cause a local disturbance in the Earth's field when it moves or stands still as shown in Fig. 5(a). The disturbance depends on the ferrous material, the size and the moving orientation of this object [81]. The disturbances are most obvious on the engine and wheels. In [82], the magnetic signature of a vehicle is described as a magnetic point dipole with a magnetic moment $m$ centered in the vehicle. It is reported that a typical vehicle has a magnetic moment of 100 ~ 300 A·m$^2$ [83]. In recent literature [82, 84, 85], triaxial spintronic sensors are employed at the center of the road lane or the roadside to detect the disturbance of magnetic field in each axis direction ($B_x$, $B_y$, and $B_z$). Here, the X-axis is parallel to the vehicle moving detection, the Y-axis is perpendicular to the vehicle moving detection, and the Z-axis is perpendicular to the road surface. The relation between the magnetic moment and the magnetic field can be described according to the Maxwell's Equations [82]:

$$B_x = \frac{\mu_0 \cdot (m_x(2x^2 - y^2 - z^2) + 3m_y xy + 3m_z xz)}{4\pi r^5}, \tag{6a}$$

$$B_y = \frac{\mu_0 \cdot (m_y(2y^2 - x^2 - z^2) + 3m_x xy + 3m_z yz)}{4\pi r^5}, \tag{6b}$$

$$B_z = \frac{\mu_0 \cdot (m_z(2z^2 - x^2 - y^2) + 3m_x xz + 3m_y yz)}{4\pi r^5}, \tag{6c}$$

where $m_x$, $m_y$, and $m_z$ are the magnetic moments in each axis direction, the $\mu_0$ is the permeability of air, and $r$ is the distance between the sensor ($x_0$, $y_0$, $z_0$) and the dipole point ($x$, $y$, $z$). The magnetic field reading in X-axis of the spintronic sensor when a saloon vehicle passes over can be briefly illustrated in Fig. 5(b) [86]. When there is no vehicle present (position A), the sensor outputs the uniform earth's magnetic field as the initial value. As the vehicle approaches the spintronic sensor, the magnetic field is distorted toward this ferrous vehicle in the negative X-axis. Then the sensor output witnesses a decrease until it reaches the minimum value (in position B). When the vehicle is exactly over the sensor, the magnetic field is slightly distorted as there is less metal in the middle position of the vehicle. As this vehicle leaves towards the positive X-axis, the magnetic field bends toward the leaving car, leading to an increase in the sensor output. After the magnetic field reaches the maximum value (in position C), it decreases until the vehicle is out of the sensing range (Position D) and then the magnetic field returns to initial value. By analyzing the disturbance signal, the presence, moving speed, direction and classification of this vehicle can be determined. To obtain a smoother magnetic field signal, a digital filtering algorithm is usually used to eliminate noise, which may utilize fast Fourier transform, median filter, and Gaussian filter, and so on.

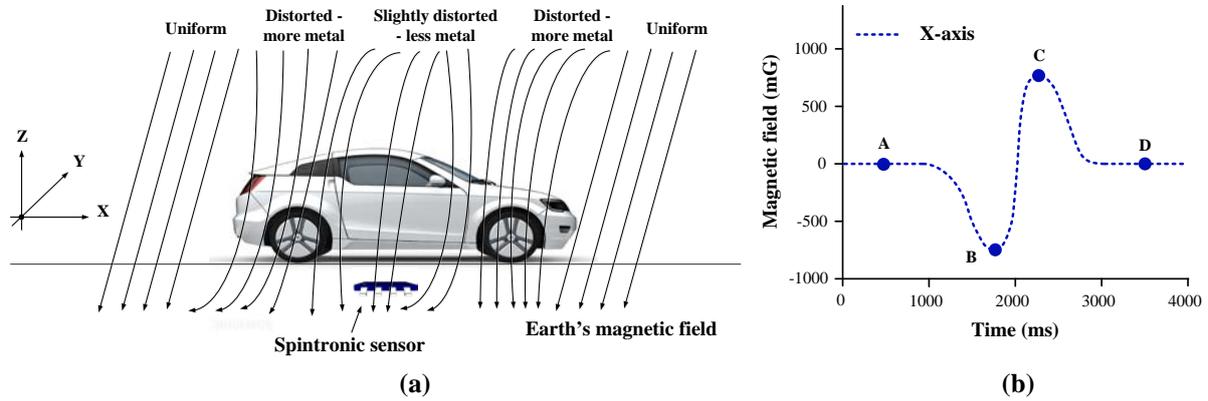

**Fig. 5** Principle of vehicle detection by magnetic field sensing. (a) The disturbance of the Earth's magnetic field by a passing vehicle. (b) The sensor reading in X-axis when the vehicle passes over the spintronic sensor [86].

The arrival and departure of a vehicle can be determined by comparing the square deviation of measured signal $SD(t)$ with the pre-defined high threshold and low threshold. The $SD(t)$ is defined as

$$SD(t) = \sqrt{(B_x - B_{x0})^2 + (B_y - B_{y0})^2 + (B_z - B_{z0})^2}, \qquad (7)$$

where $B_{x0}$, $B_{y0}$, and $B_{z0}$ are the baseline values of three axes when no vehicles pass over the sensor. The variance-based multi-state machine adaptive threshold algorithm in [87] uses the historical data variance to calculate the real-time data fluctuation, and then predict the new thresholds. This algorithm can detect the vehicle on urban streets with the precision of 97.3%. In [85] the algorithm uses the deviation of processed magnetic field intensity from a baseline to drive a fixed threshold state machine, which can detect the low-speed vehicles with an accuracy of 99.05%. Meanwhile, several other algorithms in recent literature including threshold-based algorithms, state machine algorithms, and cross-correlation based algorithms have been proposed with good results [78, 81, 88, 89].

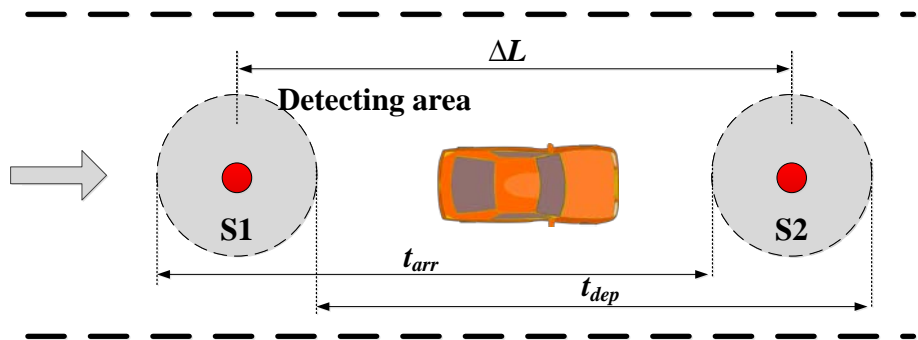

**Fig. 6** Configuration for vehicle speed estimation [82].

For the vehicle speed estimation, at least two sensor nodes (S1 and S2) are employed, as shown in Fig. 6. The distance between S1 and S2 is known as $\Delta L$, and the travel time from S1 to S2 is $\delta t$. The average speed is determined as $\bar{v} = \Delta L / \delta t$. As the time of arrival and departure points of S1 ($t_{arr\_s1}$, $t_{dep\_s1}$) and S2 ($t_{arr\_s1}$, $t_{dep\_s1}$) can be obtained, respectively, the travel time can be calculated as

$$\delta t = 0.5(t_{arr} + t_{dep}) = 0.5\left[(t_{arr\_2} - t_{arr\_1}) + (t_{dep\_2} - t_{dep\_1})\right], \tag{8}$$

In [82], the vehicle speed can also be estimated by the measurement of the signal time delay between S1 and S2 using the cross correlation algorithm.

Each category of vehicle signal has its own characteristics due to the different structures and sizes. The vehicle classification can be implemented based on a hierarchical tree methodology. Firstly the vehicle features are extracted through time-domain waveform structure after signal segmentation. The vehicle signal duration L, the signal energy E, the average energy EV, the ratio of positive and negative energy of *x*-axis VX, and the ratio of positive and negative energy of *y*-axis VY are several main parameters to classify a vehicle [85, 90]. In [85], based on a three-layer hierarchical tree model shown in Fig. 7, the detected vehicles can be classified as motorcycle, two-box car, saloon car, bus, and sport utility vehicle with the precision of 93.66% for low speed congested traffic. The threshold algorithm parameters of $L_0$, $E_a$, $EV_a$, $E_b$, $EV_b$, $VX_0$, and $VY_0$ in Fig. 8 are experimentally determined.

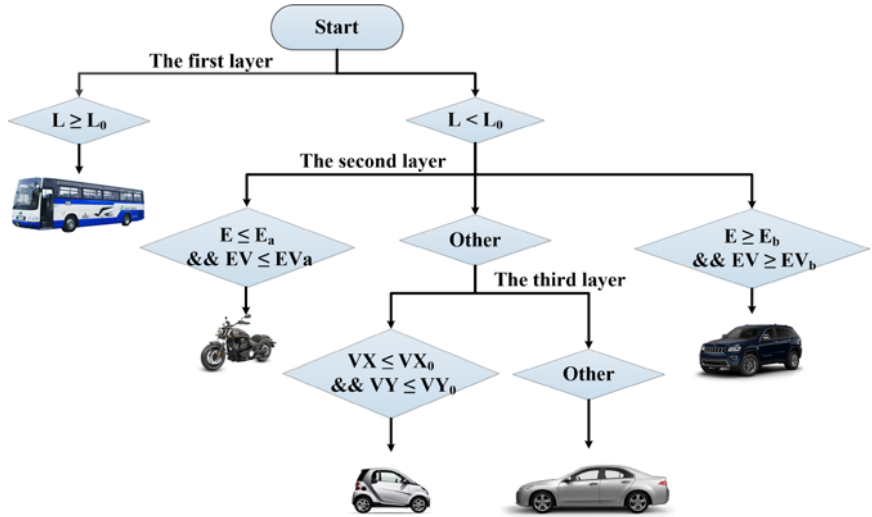

**Fig. 7** Schematic of vehicle classification algorithm [85].

In contrast to conventional detectors including inductive loops, radar sensors, and video cameras, spintronic sensors are easy to install, compact in size, low cost and immune to environment conditions such as fog, rain, and wind [84, 91]. As a result, spintronic sensors has great potential to be employed in large scale combined with the wireless sensor network for the traffic monitoring.

**2.2.4 Biodetection**

With the significant technological advancement of spintronic sensors, they have been becoming increasingly important not only in the industrial area but also in biomedical applications. Spintronic-based devices have become powerful tools for highly sensitive and rapid biological detections. The spintronic biodetection technology aims at sensing the concentration of target analyte molecules in solution, such as DNAs and proteins [92]. In the spintronic biodetection process, magnetic labels are utilized to tag the target analyte molecules, and then spintronic sensors are used to detect the magnetic signals generated from the labels [93].

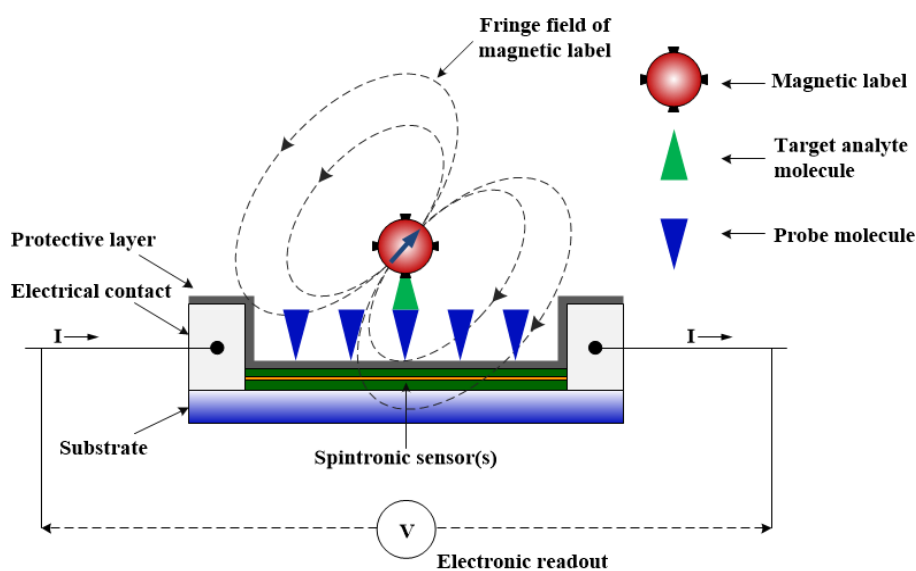

**Fig. 8** Cross-section view of biodetection using spintronic sensors [93]

The schematic of biodetection using spintronic sensors is illustrated in Fig. 8. The spintronic biodetection is implemented on the surface of a single spintronic sensor or a patterned spintronic sensor array. Firstly, the surface of the sensing area is functionalized with the probe molecules, which can specifically bind to the target analyte molecules. The target analytes in recent researches are dominantly DNAs or proteins, and the approaches to capturing them are rather different. In the DNA sequence detection, the complementary DNA chains are used as probes to capture the target DNA chain [94]. While for the detection of protein molecules, antibodies are employed to bind to the target protein through the specific immune (antigen-antibody) recognition [95]. Then the sample solution (e.g., plasma, sera, urine) is applied on the sensor surface. After the interrogation of the probe array, the target analyte molecules in the solution are specifically captured onto the sensor surface. The target analyte molecules are tagged by the magnetic labels before or after the capturing process depending on different mechanisms (direct labeling or indirect labeling) [96]. As a result, the magnetic labels are attached to the sensor surface. By exciting with the applied magnetic field, the magnetic labels produce a fringe field that is detectable by the underneath sensors. This detected magnetic signal quantitatively represents the concentration level of the target analytes in the sample solution [97].

It is noted that choosing appropriate magnetic labels is a vital aspect to the spintronic biodetection. Different magnetic labels distinctly affect the final detection signal, depending on the magnetic moment, size, structure, and surface-based binding ability [98]. Magnetic labels are normally functionalized magnetic micro-particles or nano-particles. Large magnetic particles can provide a higher magnetic moment which enables detection with a small number of particles. However, they are mismatched in size with the biomolecules and thus prejudice the quantitative capability of the biodetection. Recently magnetic nanoparticles (MNPs) with diameters under 100 nm have been widely used, as their size is comparable to the target biomolecules and are less prone to particle clustering in the applied magnetic field since they are superparamagnetic [96, 99, 100].

Conventional biodetection technology utilizes fluorescent labeling to quantify the concentration of target biomolecules. This technology requires an expensive and bulky optical system, and the sensitivity is limited [101]. Hence, it is difficult to integrate this fluorescence-based assay into a portable point-of-care (POC) device. In contrast, the spintronic biodetection technology using spintronic sensors demonstrate high sensitivity, compact size, and low cost [102]. Most importantly, the spintronic sensors provide a fully electronic readout, which can be easily integrated into electronics to detect multiple analytes on a single chip [103, 104]. Thus, this spintronic biodetection technology is a promising candidate for building portable POC devices. Actually in recent researches, several POC devices based on spintronic biodetection have been developed [105, 106]. In [107], a portable, quantitative immunoassay platform based on GMR biosensor technology can display quantitative results in less than 15 minutes with one-time user involvement, and each test costs less than US$4. In addition, among various target analytes, the protein biomarkers in plasma for early diagnosis of diseases (e.g., cancer and cardiovascular disease) have attracted much interest. These handheld biodetection devices for protein biomarkers can provide great convenience for users to realize daily diagnostic or health monitoring at the hospital, home, or outdoor environment.

### III. Winning combination: IoT and spintronic sensors

The IoT paradigm envisions to connect billions of "Things" that surround us to the Internet and expects to use the information of these "Things" to enhance utilization effectiveness and efficiency of various public resources, thus accelerating the actualization of smart living. Spintronic sensors, as introduced in the previous section, can provide the information of magnetic field and magnetic-field-related parameters (e.g., current, vehicle speed, analyte concentration, etc.) in several applications, which can be a cornerstone for the IoT. More specifically, wireless spintronic sensor networks (WSSNs), which will be discussed in this section, can seamlessly integrate spintronic sensors into the IoT platform. This combination of the IoT and spintronic sensors has great potential to tackle roadblocking challenges in the fields of building, power grid, transport, and healthcare. In this section, we present a brief introduction of the IoT concept and architecture, and then discuss the proposal of the WSSN. It should be noted that it is not our purpose to provide a comprehensive survey of the IoT in this section. We aim at emphasizing how the spintronic sensors can work with the IoT platform.

### 3.1 Brief introduction to the IoT

Since the term of IoT was first coined by Kevin Ashton in 1999 [108], a number of definitions of the IoT have been proposed. In [109], the IoT can be defined from three visions: internet-oriented, things-oriented, and semantic-oriented, derived from the perspectives of middleware, sensors, and knowledge, respectively. Actually, the IoT paradigm is the result of the convergence of these three visions. It is commonly accepted that IoT is a global infrastructure where day-to-day digitally augmented objects can be equipped with the capabilities of sensing, identifying, processing and networking, then they can communicate with other devices over the Internet, and finally accomplish specialized objectives [110].

Standardizing a common architecture for the IoT is a very complex task as billions or trillions of heterogeneous objects, various link layer technologies, and diverse services may be involved in the IoT system and many factors like security, privacy, reliability, scalability, interoperability, QoS (Quality of Service), etc. should be considered. To build such a common architecture, manifold architectures frameworks for the IoT have been attempted in recent literature. At the initial development stage, the three-layer architecture composed of the perception layer, network layer, and application layer is widely adopted [111, 112]. However, this IoT architecture lacks effective management methods and business models. Afterward, to obtain a more reasonable architecture for IoT, several architectures including the middleware-based, SOA (Service Oriented Architecture) based, and five-layer architecture have been proposed [113, 114]. Among these architectures, the five-layer architecture is established by combining the TCP/IP (Transmission Control Protocol/Internet Protocol) model, the TMN (Telecommunications Management Network) model, and features of the IoT, which has been utilized in [111, 113]. This IoT architecture can be illustrated in Fig. 9, and each layer is briefly described as follows:

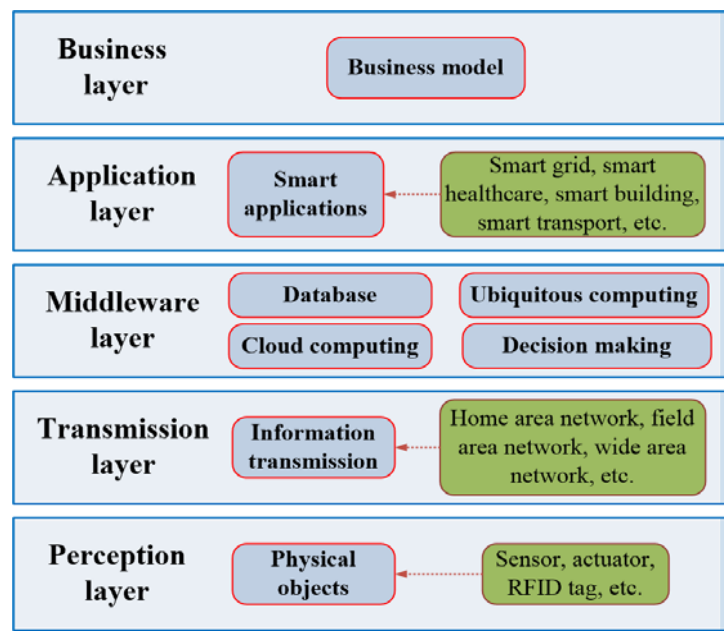

**Fig. 9** The five-layer IoT architecture [114]

a) Perception layer represents the physical objects, such as sensors, actuators, and RFID tag, which aim to recognize things, collect information of interest, and convert this information into digital data. Depending on the type of sensors, the collected information can be temperature, humidity, location, current, etc. In this paper, we focus on the spintronic sensors, which collect the information of magnetic field and magnetic-field-related parameters. The digitalized signal is then passed to the transmission layer through secure channels.

b) Transmission layer (or called network layer) securely transfers the data passed from the perception layer to the middleware layer through various networking technologies. The primary communication technologies can be classified as home area networks (HAN), field area network (FAN), and wide area networks (WAN), depending on the practical requirements of location, data rate and coverage range [115, 116]. An overview of communication technologies for IoT applications is shown in Table 1.

Table 1. Common communication technologies for the IoT applications

| Networks | Technology | Typical Frequency | Range | Data Rate | Unique Features | Reference |
|---|---|---|---|---|---|---|
| HAN | RFID | 13.56 MHz | < 1m | 6.6 ~ 26.5 Kbps | Possibility of unique IDs; high speed data capture; simultaneous multi-tag reading; | [117] |
| | NFC | 13.56 MHz | < 10 cm | Up to 424 Kbps | Peer-to-peer communication; short set-up time; | [118] |
| | ZigBee | 868/915/2400 MHz | 50 ~ 300 m | 20 Kbps, 40 Kbps, 250 Kbps | Self-deployment; support of addressing and routing for the Ad hoc, tree and mesh topologies, providing flexible network structure; | [119, 120] |
| | Z-Wave | 868/908/2400 MHz | 30 ~ 100 m | 9.6 Kbps, 40 Kbps, 200 Kbps | Low powered RF communication; support of full mesh networks; | [120] |
| | Bluetooth | 2.4/5 GHz | 10 ~ 100 m | 1 Mbps, 24 Mbps | Complicated set-up procedure; Ad hoc topology; | [121] |
| | Wi-Fi | 2.4/5 GHz | < 300 m | < 300 Mbps | The most widely used WLAN technology; high data rate; | [122] |
| | Dash7 | 433/868/915 MHz | < 5 km | < 10 Kbps | Low energy standby mode; support of peer-to-peer, star, and mesh topologies; | [115, 123] |
| FAN | PLC (Power line communication) | 1 ~ 30 MHz | 1 ~ 3 km | 2 ~ 3 Mbps | Using existing power cables to simultaneously carry both data and electricity current; | [124] |
| WAN | Cellular (GSM, GPRS, 3G, LTE) | 700 MHz ~ 2.7 GHz | < 50 km | < 300 Mbps | Existing communication infrastructure; widespread and cost-effective; | [124, 125] |
| | WiMAX | 2.5/3.5/5.8 GHz | < 50 km | < 75 Mbps | Advanced IP-based architecture; flexible channel bandwidth to facilitate long range transmission; | [126] |

c) Middleware layer is a software layer interposed between the transmission layer and application layer, which can directly match services with the corresponding requesters and has a link to the database. The database, ubiquitous computing, cloud computing and decision making can take place in this layer [111]. For instance, in the scenario of smart transport, the middleware layer can process the received data about the traffic information (e.g., traffic flow, occupancy rate, etc.) to predict the future traffic conditions.

d) Application layer provides various high-quality services and applications to users or customers according to the processed data in the middleware layer. Smart applications such as smart building, smart grid, smart transport, and smart healthcare (which are discussed in Section 4) are implemented in this layer. This layer also can provide the interface for customers to interact with a physical device or to access the designated data.

e) Business layer at the highest level is capable of managing the overall system. Based on the received data from the application layer, this layer can build several business models. It is noted that the success of a technology depends on both the advances in technology and the sound business models. A successful IoT business care is Uber [127], a ride sharing, car-hailing app that taps into the data enabled by IoT to deliver service to both drivers and passengers. In addition, this layer should also manage the research on the privacy and security, which are the challenging elements in the development of IoT. For example, this layer only provides the access for the authorized users.

This multiple layered architecture of the IoT clearly indicates that the realization of the IoT relies on the integration of multiple enabling elements. These IoT elements can be categorized as hardware (e.g., identification, sensors, actuators, communication), middleware (data storage and analytics), and presentation (which presents meaningful information and services to the end-users) [128, 129]. Among these enabling elements, wireless sensor network (WSN) [130] plays an especially important role as hardware to support the IoT. A WSN is a network formed by extensive specialized sensor nodes with a communication infrastructure, which monitors physical or environmental conditions at diverse locations and cooperatively transfers the collected data through the network to a main location. One way to realize the IoT is connecting WSNs to the Internet. Nowadays the WSN-based IoT platform is the popular solution in remote monitoring and management applications. In this paper, we propose a WSSN which mainly utilizes the spintronic sensor for the collection of information of interest.

**3.2 Wireless spintronic sensor network (WSSN)**

A common WSN-based monitoring network is mainly composed of four components [128]: (a) Hardware - A typical wireless sensor node contains sensors, processing units, transceiver units, power source, etc. Commonly the sensor nodes are of small size, light weight, and low power consumption. (b) Communication stack - WSN nodes communicate among themselves to transmit data in a single or multi-hop fashion to a base station. (c) Middleware - a platform-independent middleware can combine cyberinfrastructure with a Service Oriented Architecture (SOA) and sensor networks to provide access to heterogeneous sensor resources. (d) Data aggregation - an efficient and secure data aggregation method can extend the lifetime of the network and ensure reliable data collected from sensors. Energy efficiency, scalability, reliability, and robustness should be considered when designing a WSN solution [109]. In addition, the recent advances in lower-power integrated circuits, embedded processing, and wireless communication technologies have driven more efficient and stable wireless sensor network.

As described in Section 2, spintronic sensors are compact in size, low cost, low energy consuming, and highly robust. As a result, spintronic sensors can meet the requirements of smart sensors in a WSN. Most importantly, the WSSNs are possible in the large-scale applications.

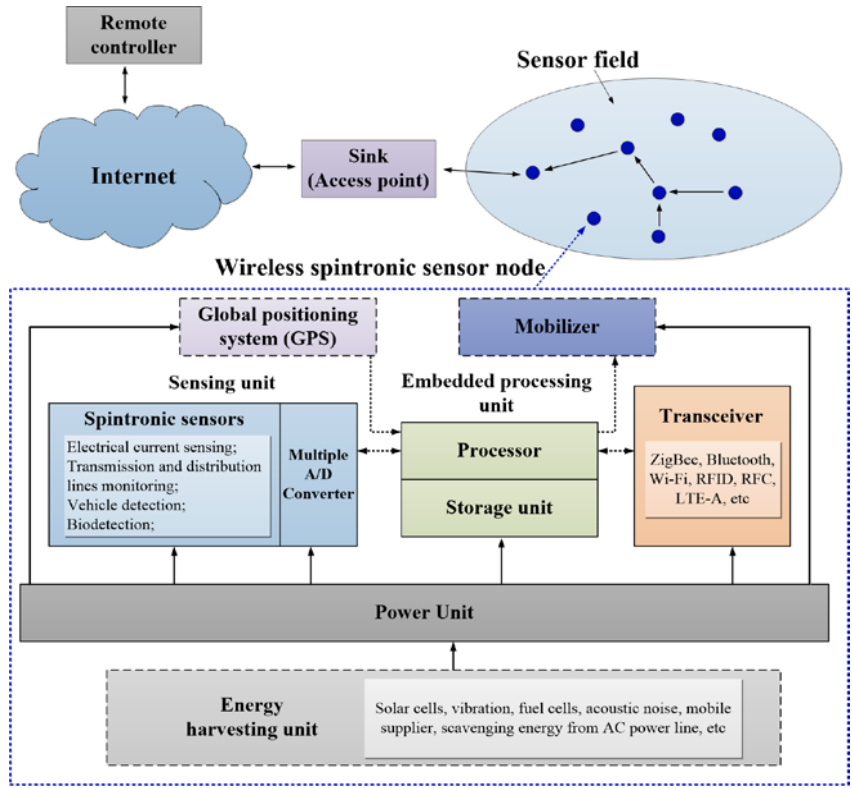

**Fig. 10** Hardware structure of a wireless spintronic sensor node [131]

The hardware structure of a wireless spintronic sensor node is illustrated in Fig. 10. A standard wireless spintronic sensor node is composed of four essential components: a spintronic sensor unit, an embedded processing unit, a transceiver unit, and a power unit. Depending on the practical applications, additional components such as global positioning system (GPS), energy harvesting unit, and mobilizer are also required [130]. A spintronic sensor unit consists of spintronic sensors and multiple A/D converters. The spintronic sensors can sense the magnetic field signals in the applications, and then these analog signals are converted to digital data by the A/D converters. The microprocessor in the embedded processing unit can process the digital data and arrange this sensor node to collaborate with other nodes. Generally, this processing unit associates with a storage unit. Nodes report the detected information to a small number of special nodes called sinks (access points). Here, the transceiver unit can connect this node to the sinks through various wireless communication technologies (see Table 1). Most importantly, a power unit, commonly battery power, provides the energy of all the system. It should ensure a sufficiently long lifetime of the sensor node. Moreover, the power unit can be supported by an energy harvesting unit that scavenges energy from the external sources, such as solar cells, fuel cells, vibration, acoustic noise, a mobile supplier, and AC power line [130, 132, 133]. Certain remote monitoring tasks (e.g., in power grid and transport monitoring systems) require the information of position at high accuracy. Hence, a GPS within at least 5 meters of accuracy is commonly deployed within the sensor node [134]. Furthermore, sometimes a mobilizer may be needed to move the sensor nodes in some particular tasks (e.g., changing the antenna's orientation [135]).

A bunch of sensor nodes can constitute an interconnected WSSN. Combining the superb measuring capabilities of spintronic sensors and the pervasive monitoring ability of WSNs, the WSSNs have high potentials in the monitoring applications which are related to the measurement of magnetic field. In addition, spintronic sensors and other types of sensors can be integrated into a wireless sensor node, working collaboratively to accomplish the measurement of several special parameters, e. g. a spintronic sensor measuring electric current can be integrated with a voltage sensor to measure the power consumption of an electric appliance.

## IV. IoT and spintronic sensors enabling smart living

In the previous section, we discussed the WSSN that integrates spintronic sensors with IoT technology. This section presents four different scenarios in which the WSSN-based solution can constitute pervasive monitoring systems: smart building, smart grid, smart transport, and smart healthcare. Such monitoring systems can collect large-scale system information of interest and permit dynamic updates for the administrators and end-users. In this way, several roadblocking challenges to actualize smart living can be tackled, which cannot be achieved by the traditional monitoring systems.

### 4.1 Smart building – pervasive building energy management system

Buildings are the largest energy-consuming sector, which contribute to about 40% of the global energy consumption and 30% $CO_2$ emissions [136]. One major challenge in smart building is reducing the overall energy consumption and carbon footprint of buildings without compromising comfort. It is estimated that up to 30% of a building's overall energy consumption can be saved by optimized operations and managements without changing the structure or hardware configuration of the building [137], among which reducing the wasteful usage of energy and implementing demand response (DR) are the feasible approaches to help tackle this challenge. On one hand, in a residential or commercial building, approximately one-third of the overall energy is consumed by various electric equipment such as HVAC (heating, ventilation, and air conditioning) systems, lighting, elevators, and other electrical appliances [138]. The wasteful usage of such building equipment/devices causes a large portion of energy waste due to occupants' unawareness, stand-by consumptions, and non-adaptive control of electric appliances. For example, electric appliances in standby mode waste approximately 10% of the total energy in a buildings [139]. A large part of energy is consumed due to the wasteful usage of HVAC and lighting systems without adaptive control. However, the traditional way of measuring the overall energy consumption of buildings cannot detect the wasteful usage of individual building appliances. Even through smart meters can dynamically monitor the overall energy consumption of a household, they cannot provide the specific information about the energy consumption of individual electric equipment/devices. It is crucial to monitor real-time energy consumption of each appliance to recognize the possible wasteful usage, and remotely control the behavior of appliances accordingly. On the other hand, DR is a scheme where the end-user energy consumption patterns change in response to the real-time price of electricity or incentive payments designed to shift their load from peak to off-peak periods in order to avoid the situation that the power system balance is jeopardized [140]. It can be an enabler for renewable energy which can fluctuate with variable weather conditions, driving down $CO_2$ emission.

DR heavily relies on large-scale intelligent electricity meters and advanced communication technologies. The conventional electricity meters do not provide real-time tariff information for DR scheme, and thus the customers cannot make consumption decisions on a daily or hourly basis. Although there is existing smart meter technology, normally only one smart meter is installed in a household because of its inconvenient installation that is invasive and interrupts the power lines. Here, a smart energy management system for buildings (see Fig. 11) enabled by pervasive spintronic power meters and IoT technologies can help achieve the above two approaches. This system is composed of two main components: a monitoring subsystem and a control subsystem, and they are based on the wireless sensor and actuator networks.

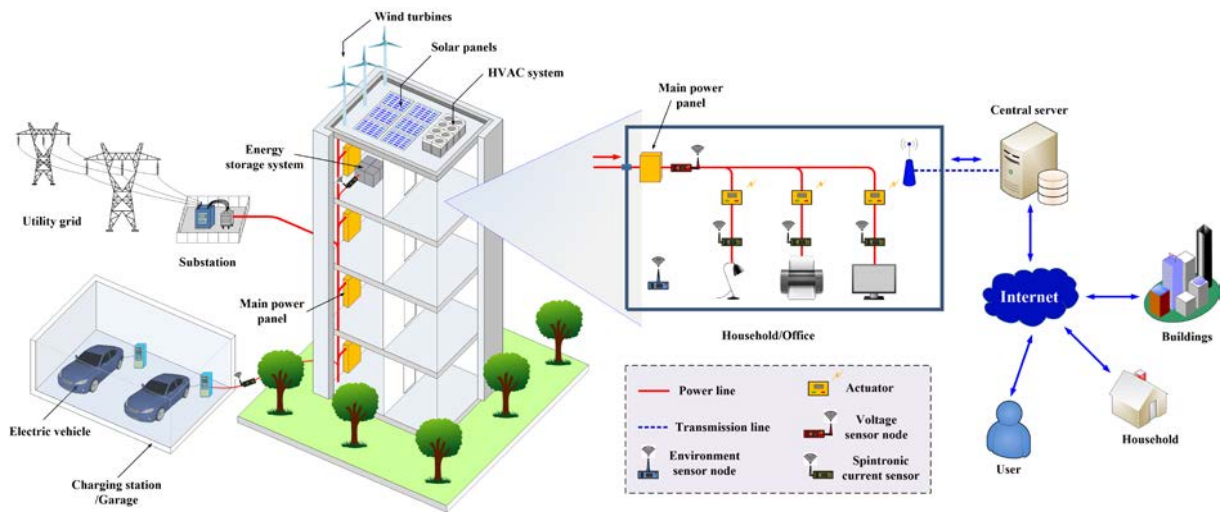

**Fig. 11** The architecture of the pervasive energy monitoring system for smart building [141]

In the monitoring subsystem, the set of spintronic power meters in each household or each floor forms a WSN. The power meter consisting of both spintronic sensors and voltage sensors is employed through the power lines to measure the energy consumption/generation of each electric appliance or equipment [142]. As mentioned in Section 2.2.1, spintronic sensors can sense the currents in multi-core conductors (commonly two-wire power cord) without clamping, which can conveniently measure the working current $I(t)$ of residential/office appliances or equipment (e.g., lamp, air conditioner, TV, printers, electric vehicles, etc.) or other electric generating component (e.g., solar panel, wind turbine) in a non-contact manner. Meanwhile, non-contact voltage sensors [143, 144] can measure the working voltage $U(t)$ by detecting the electric field nearby the power bus. Thus, the instantaneous power $P(t)$ of an appliance is determined as $P(t) = I(t) \times U(t)$. Through sampling the power of this appliance in real-time, the energy consumption is calculated as $E = \int P(t)dt$. The temperature, humidity, lighting level and air quality are the primary parameters affecting comfort level [136, 145]. Hence, environmental sensors that measure these parameters should also be deployed simultaneously in each household to adapt the behavior of the HVAC and lighting system. The data collected by spintronic power meters and environmental sensors can be wirelessly transmitted to the WSN access point located in the same household or floor, and then conveyed to a central server via a transmission line interface. The central server has the responsibility to perform the aggregation, process, and storage of the real-time and historical data. By virtue of the specialized user interface (e.g., RESTful Application Programming Interface [146]) and the Internet, users can access the power consumption/generation and status (i.e., on/off/standby) of each appliance through a web browser on smartphones, PDAs, laptops, and PCs [147]. The information enables the users to clearly recognize the wasteful usage of appliances. It is noted that providing informed feedbacks to the end-users can reduce the overall energy consumption in the order of 5% ~ 20% [148]. The advantages of spintronic sensors such as low cost, compactness, low energy consumption, non-contact measurement, and easy installation enable spintronic power meters to be deployed in building energy management system on a large scale. On the other hand, the control subsystem aims at controlling the operation of electric appliances (e.g., switching on/off or regulating the current of appliances) based on the provided information from the monitoring subsystem and the specific energy conservation strategies [139]. The control infrastructure possesses a large number of actuators for individual appliances. These actuators wirelessly receive the control commands from the central server and take the corresponding actions. For instance, the control system will switch off an electric appliance after a predefined time interval in standby mode.

Furthermore, such smart building energy management system can significantly facilitate DR strategy programs with the aid of the pervasive spintronic power meters and the IoT platform. The pervasive spintronic power meters combining advanced IoT technology can be widely deployed in buildings or households to measure and record energy consumption data on demand side at hourly interval or more frequently (e.g., nearly real-time), and provide usage data to both customers and electricity companies [149]. The customers can then schedule the tasks of appliances at an appropriate time. For instance, several white goods (e.g., washing machine, dishwasher) can be scheduled to execute the task when energy cost is lower (e.g., at night) [150]. The historical information of building power consumption (e.g. 1-year-long building energy usage) can provide the weekly, monthly and seasonal patterns, which enables the forecasting of the short-term future energy demands, enabling companies to offer up-to-date tariff options such as time-of-use (TOU) rates, critical peak pricing (CPP), and real-time pricing (RTP), which will encourage consumers to migrate their electricity usage to off-peak period [151]. On the other hand, renewable energy generation such as solar panel and wind turbines over the rooftop [152], energy storage systems (ESSs) [153] and vehicle to grid (V2G) [154] can be integrated into a building to achieve DR by constituting a microgrid [155]. Pervasive spintronic power meters can monitor the energy generation and/or consumption of these components in real-time, which contributes to autonomous DR optimization by load controller and building energy management systems.

Overall, this intelligent building energy management system enabled by pervasive spintronic power meters (and possibly working together with other environmental sensors) and IoT technology ensures reduction of energy consumption and facilitating DR, and at the same time potentially enhances the comfort level perceived by the end-users, therefore providing a more sustainable and smart building.

## 4.2 Smart grid – wide-area transmission and distribution network monitoring system

Smart grid, first officially coined in the Energy Independence and Security Act 2007 of America [156], is to modernize the transmission and distribution grids for a more reliable and secure infrastructure that meets the future demand and achieves intelligent control. Integration of renewable energy and self-healing are the key challenges to be tackled by smart grid, which cannot be effectively implemented by the conventional power grid system. The monitoring system in the conventional power grid is established in a centralized way (i.e., with monitoring sensors mainly deployed in substations), and cannot effectively fulfill these two tasks. Both integration of renewables and self-healing in power grid need more dynamic and pervasive monitoring information than that the existing power grid system provides.

Multiple renewable energy such as solar energy, wind energy, tidal energy and geothermal energy are under development over the world to replace the traditional fossil energy. Scientists have advanced a plan to power 100% of the world's energy with wind, hydroelectric and solar power by 2030 [157]. However, our existing transmission grid was not designed to carry the additional capacity of renewable energy on a large scale. Meanwhile, it is very expensive to build new transmission lines (US$ 1 million/mile), and also it is hard to get all approval from the government, local communities, and environmentalists. One possible approach called dynamic line rating (DLR) [158] is to dynamically stretch the limit and overload the existing transmission lines to carry the renewable energy. The rating of the transmission lines is found to increase around 20 ~ 30% by DLR at most of time [159]. However, the extra current loading can overheat the conductors to exceed the design temperature which can cause detrimental annealing and damage the cable. Moreover, the heat expansion of the conductors can lead to the lowering in the cable height particularly at the mid-span section, a phenomenon called sagging [160]. The clearance between the conductor and the surface may become too small, violating the safety clearances and resulting in possible short circuit between the conductors and the ground. DLR needs a real-time monitoring technique to keep track of the spatial positions of overhead transmission lines, and transmits these dynamic information to the Supervisory Control and Data Acquisition (SCADA) system to enable the optimization strategies for transmitting more power without violating the safety margin. On the other hand, the residential, commercial and industrial end-users are demanding more stable and sustainable power supply, since the power failure can bring human life into chaos and result in huge economic losses. Pinpointing the fault location as quickly as possible is critical for re-configuring the power network for self-healing [161]. It is traditionally very hard to isolate the problematic part of the network accurately in a timely manner because the fault point needs to be searched over the whole span of the transmission line by analyzing the voltage and current recorded at the busbar when the fault occurs. Meanwhile, the difference between the calculated and real faulted point can be very large even with a small error in measurement. The self-healing grid under development is promising for isolating the faulted area within the minimum range by reconfiguring the switches and reclosers installed on the distribution feeder and re-establish services to as many customers as possible from alternate sources/feeders [162].

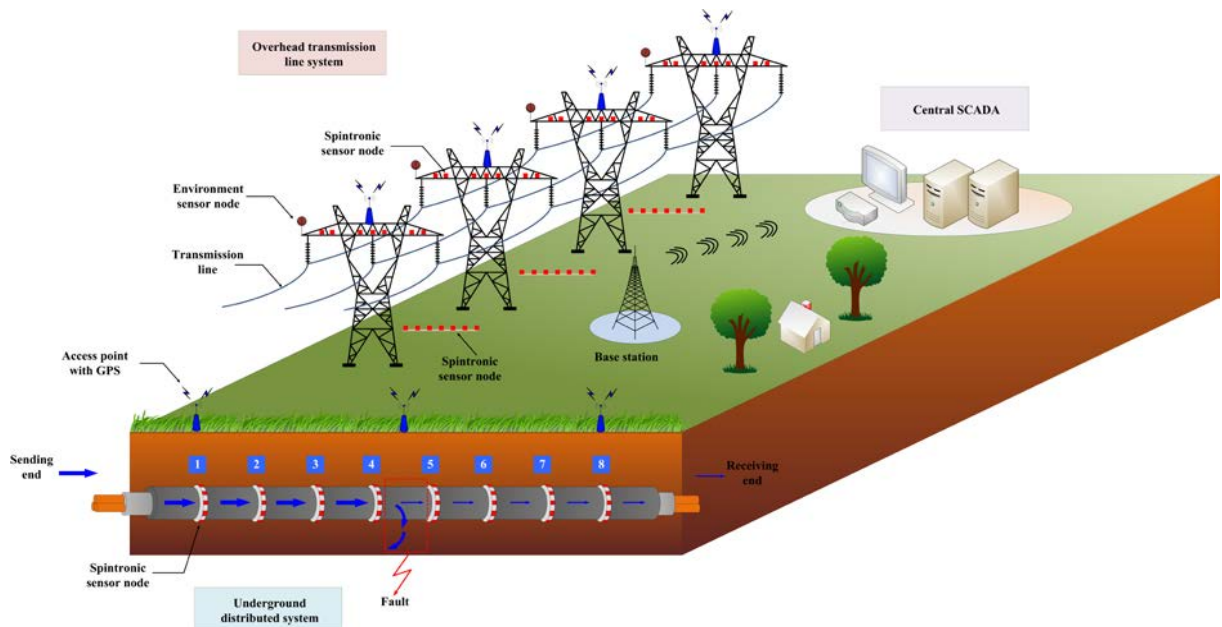

**Fig. 12** Schematic diagram of wide-area transmission and distribution lines monitoring system by deploying spintronic sensors with IoT technology.

The developed WSSN combined with IoT technology is a promising candidate to solve the above challenges by constructing a wide-area transmission and distribution network monitoring system, which is illustrated in Fig. 12. As described in Section 2.2.2, spintronic sensors can be deployed on a large scale for wide-area monitoring of transmission and distribution network. For the overhead transmission lines, the sagging conditions and the operating currents can be acquired from spintronic sensors installed on the transmission towers, combining with real-time environmental information measured by a variety of environmental sensors measuring the physical conditions of transmission lines (e.g., temperature, humidity, ambient wind speed, and solar radiation). All these sensor nodes sense the magnetic field and environmental conditions, and periodically transmit the data to the access points using PLC or other short-range radio communications (see Table 1). Then wide area network transfer it to the central SCADA. The real-time maximum instantaneous current carrying capability can be analyzed to make sure operating within the safety limit.

On the other hand, together with the spintronic sensor arrays distributed along the underground cables, the search for the faulted point can be narrowed from adjacent substations to adjacent towers for the overhead lines (or adjacent sensor arrays for the underground power cables). For the overhead transmission lines, the fault point can be estimated by analyzing the magnetic fields emanated from the transmission lines [163]. For the underground distribution cables, the magnetic field patterns measured by the sensor array mounted around the cable surfaces are different when the fault happens. All these information is very useful for fault location and inferring reason resulting in the incident. The workload of pinpointing the fault location is significantly narrowed. As such, this wide-area transmission line and distribution network monitoring system taking advantage of spintronic sensors will enable the DLR and rapid fault location, thus effectively achieve the integration of renewable energy and self-healing grid, respectively. It also provides better situational awareness for the operation staff to react and make predictive decisions for ensuring the reliability of the power grid. All these developments in smart grid present an unprecedented opportunity for developing small living to change the way we live and work.

**4.3 Smart transport – all-round traffic monitoring and management system**

Nowadays the increasing population and growing size of cities lead to a rapid increase in the amount of vehicles on the roads. One of the most critical consequence is the management problem of traffic congestion. Traffic congestion causes huge economic losses worldwide, e.g., $101 billion for USA and €200 billion for Europe in 2012 [164]. Hence, effective traffic congestion management is crucial as it affects both public and private vehicles and also adversely affects the environment. The existing traffic management system lacks the ability of granular data collection, and cannot provide timely and sufficient traffic information for highly-efficient traffic management. An intelligent traffic management system is expected to reduce traffic congestion and improve response time to the accidents, ensuring a comfortable travel experience for commuters. Here, a WSSN-based traffic monitoring system that integrates spintronic sensors and IoT technologies providing detailed dynamic traffic information and consequently facilitating the highly-efficient management of road traffic is illustrated.

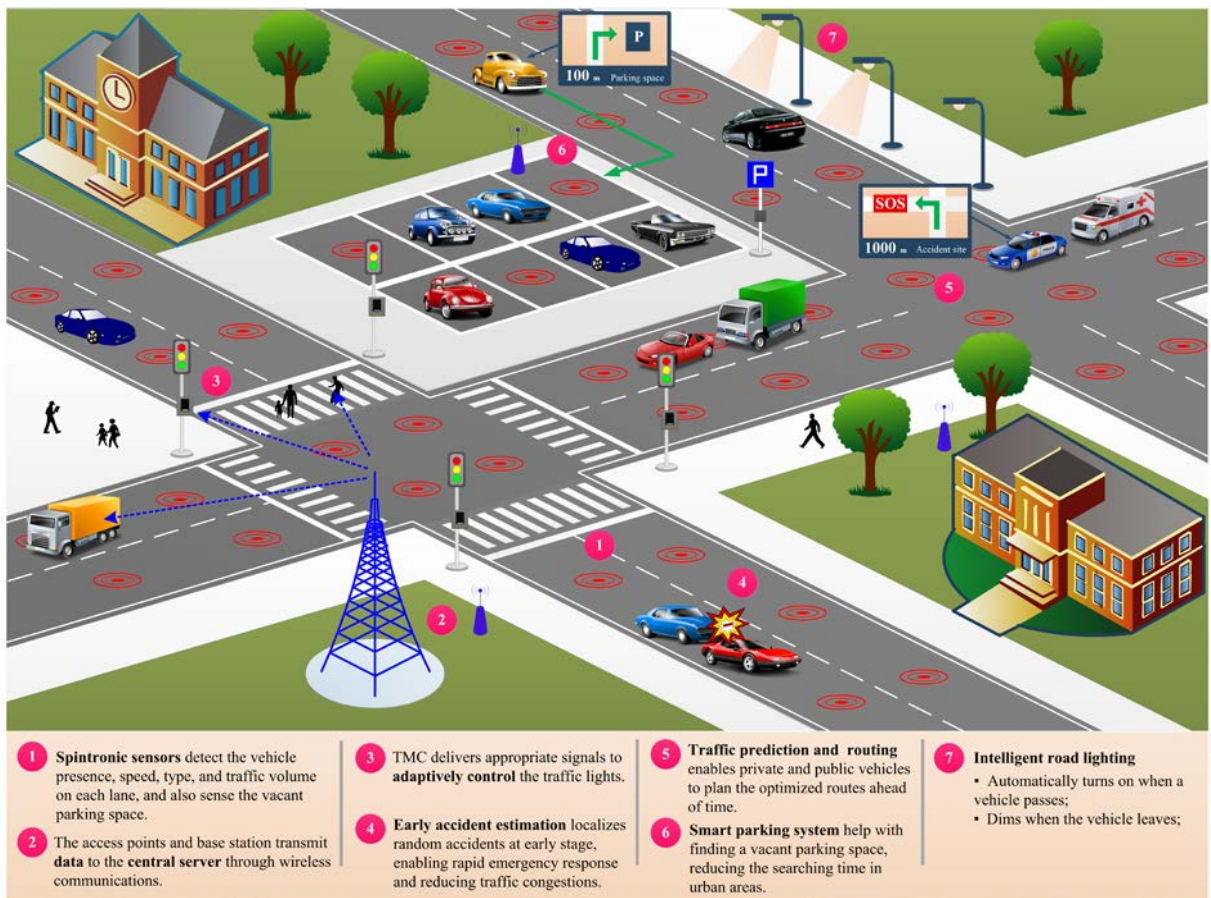

**Fig. 13** System architecture of the WSSN-based traffic monitoring and management system [164, 165]

A schematic architecture of the WSSN-based traffic management system is illustrated in Fig. 13. In this scheme, high density of wireless spintronic sensor nodes are installed on the surface or at the sides of the roads. As described in section 2.2.3, spintronic sensors can accurately measure the dynamic traffic information such as vehicle presence, speed, type, occupancy rate, and traffic volume on each lane. Through the short-range radio communications (see Table 1), each sensor node containing GPS reports the traffic data to the access point is normally located at the roadside. The access point then transmits the data packet to a remote server via cellular network communication. After the data aggregation, process, and storage by the server, the processed traffic information ultimately can be accessed by the traffic management center (TMC), users, and vehicles. Additionally, the vehicle-to-infrastructure (V2I) communications also enable the sensing nodes to deliver timely traffic information directly to nearby vehicles [166].

In conventional traffic monitoring system, only a small number of sensing devices such as video cameras, radar sensors, and inductive coils are deployed in the road intersections due to the high cost of installation and maintenance. In contrast, low-cost and compact wireless spintronic sensor nodes can be densely installed along the roads and at intersections (e.g., with inter-distances ranging from 0.5 ~ 2 km [165]), which ensures a comprehensive and higher-resolution sampling of traffic dynamics for the ITS applications to enhance traffic efficiency and safety. Firstly, the real-time estimation of the number and type of vehicles approaching a road interaction can be achieved, which permits TMC to deliver appropriate commands to traffic control unit for the adaptive control of traffic lights [115]. Such an adaptive traffic light control will replace the conventional approach with fixed switch interval times, and achieve more balanced traffic distribution by minimizing delays and unnecessary stoppages for vehicles, efficiently reducing the traffic congestion at intersections. Secondly, the short-time traffic evolution can be predicted according to the historical traffic data, the real-time traffic conditions and the forecasting weather conditions [164]. It is vital to predict the dynamic traffic flow during a large-scale event (e.g., live shows, sports events) or commuter traffic in huge urban spaces. The real-time traffic monitoring system provides a dynamic route planning for the drivers at early stage. For example, the TMC provides users in advance with the precise and valid traffic conditions; then the drivers can ahead make an informed decision to optimize their routes, thereby avoiding traffic congestion and reducing their journey time and energy consumption [167]. Thirdly, this system can detect and locate the random traffic incidents and immediately report the incident time and location to the TMC. Hence, the road traffic prediction enables faster and more efficient emergency services dispatch (i.e., ambulance, police, and fire engine) to the incident points in optimized routes, which greatly reduces the response time [168]. According to the U.S. Federal Highway Administration, over 25% of traffic congestion is caused by traffic incidents [169]. Early incident detection and emergency response contribute to less congestion and safer roads. Fourthly, advanced notifications of traffic congestion ahead, obstacles on the road (e.g., broken vehicles), and incoming vehicles can be provided for the drivers and pedestrians. Such notifications will encourage safe and defensive driving, reducing the rear-end collisions with vehicles that are stopped or traveling at low speed and other crash accidents with incoming vehicles that are making a turn at interactions [170]. Meanwhile, warning and alerts can be provided for pedestrians in the crosswalk at the interactions or school zones when the vehicles are approaching the crosswalk. Such measures can reduce the number of accidents for the vulnerable groups on the roads. Fifthly, variable speed limits can be adjusted on motorways depending on the real-time traffic volumes, which can improve the traffic flow, reduce the number of accidents and alleviate traffic congestion [171, 172].

The WSSN-based intelligent traffic management system also promotes other ITS applications. A smart parking system can be achieved with the aid of the WSSN-based monitoring system [79, 173]. In this system, the WSSNs detect the occupancy of parking space and then provide the real-time information to the drivers. For the drivers who have booked parking spaces in advance using smartphone applications, the system can guide the drivers to the vacant spaces. Nearly 30% of drivers in cities are searching a parking space, and 4.5 km is the extra distance that drivers have to travel on average to find a parking spot [174]. Such a smart parking system enables the drivers to save time to find a parking space in urban areas (especially in the central business district) and results in less traffic congestion. RFID technology can also be used here for recognition. Similarly, the WSSN can detect the available charging spaces in a large charging station for electric vehicles (EVs). The EV drivers can be informed of the nearest locations of vacant charging space.

In addition, vehicle detection can be applied to control the road lighting adaptively in spare suburban roads at night during low traffic period. With the help of dense wireless spintronic sensor nodes, the lighting system automatically increases the appropriate light intensity before the vehicle approaches and dims it out after the vehicle leaves. Such an intelligent road lighting system can achieve over 90% energy saving [175], leading to a green transport system. Besides, the city traffic data will be useful for bus management systems to design the bus routes and optimize the operation of the bus fleets in order to provide most efficient bus service [176]. Another interesting application using dense wireless spintronic sensor nodes in intelligent traffic system is train detection [177, 178]. The WSSN can detect the occupancy, speed, length of a moving train on the rail, enhancing the safety and reliability of high-speed rail system. For instance, such a system can avoid the train collision by ensuring that no more than one train enter a rail section at the same time.

The strong data-driven analytical capability and automated traffic pattern visualization to be enabled by the system can enhance the area-wide optimization strategies and potentially transforms government, business, and society. City government traditionally rely on the static statistic data or expensive, custom on-site surveys to visualize the traffic movement in cities; they can derive insights from multiple network data in real-time to improve traffic planning and manage highway toll pricing and parking space better [179]. From the perspective of city planning, the historical or real-time traffic information can be utilized to analyze the traffic patterns for traffic infrastructure and policy planning in the future. For instance, city planners will be able to make more informed decisions about where to place new bus stops and parking lots, how to add racks and widen lanes, and even how much road salt to apply after a heavy snow. Such smart transport system is a necessity for the success of other city sectors and the creation of jobs, and will play a key role in cultivating productivity and invigorating economy. On the whole, the WSSN based traffic management systems will slowly revamp the modern metropolis by providing more advanced traffic infrastructure and services, and the citizens can enjoy a smarter and greener lifestyle.

**4.4 Smart healthcare – pervasive health diagnosis system**

One of the major challenges of the world in the recent decades has been the continuous population increase, especially the rise of the elderly populations. The need of delivering quality healthcare to a rapidly growing population while reducing the healthcare costs in modern-day society is a challenging issue [180]. In the existing healthcare systems, patients are monitored or diagnosed in the premises of healthcare. Meanwhile, only limited quality healthcare services can be provided for the growing populations. Moreover, the growing healthcare cost is a universal concern. An enabling component of affordable global healthcare is the pervasive health diagnosis systems with the help of point-of-care (POC) healthcare technologies and the advanced IoT platform. POC technologies are considered as an effective means of reducing healthcare costs and improving efficiency, which are critical in the provision of diagnostic and monitoring healthcare in countries with large populations or rural areas. Such a pervasive diagnosis system requires the combination of smart portable bio-sensors, computing, networking, and information and communication technologies. Here, a feasible pervasive diagnosis system based on spintronic biodetection and cloud services is illustrated. Fig. 14 shows a schematic framework of this pervasive diagnosis system which consists of four basic parts: smart POC devices, data transmission, cloud services, and multiple users [181]. As described in section 2.2.4, one POC device enabled by spintronic biodection technology comprises single or multiple spintronic sensors, and can detect the versatile target biomolecules, predominately DNA and protein in present researches. Up to now, the research groups from the USA [106] and Portugal [102] have built prototypes of POC device by miniaturizing the spintronic biodetection platform into a handheld and battery powered device, and the detection of protein [106] and DNA [102] was realized. These low-cost and highly portable POC devices can be widely operated in the hospital and at home for users' personalized health management, or even in the outdoor environment as part of the emergency medical services (EMS). The test data from POC devices is then transmitted to the "cloud" through wireless communication and the Internet. More importantly, the hybrid cloud composed of public cloud and private cloud can make full use of the abundant monitored data from the pervasive POC devices. Multiple cloud services including automatic diagnosis and medical decision making can be rapidly provided. Different end-users such as patients, clinicians, and hospitals can ubiquitously obtain these cloud services through user interfaces in their PCs or smartphones.

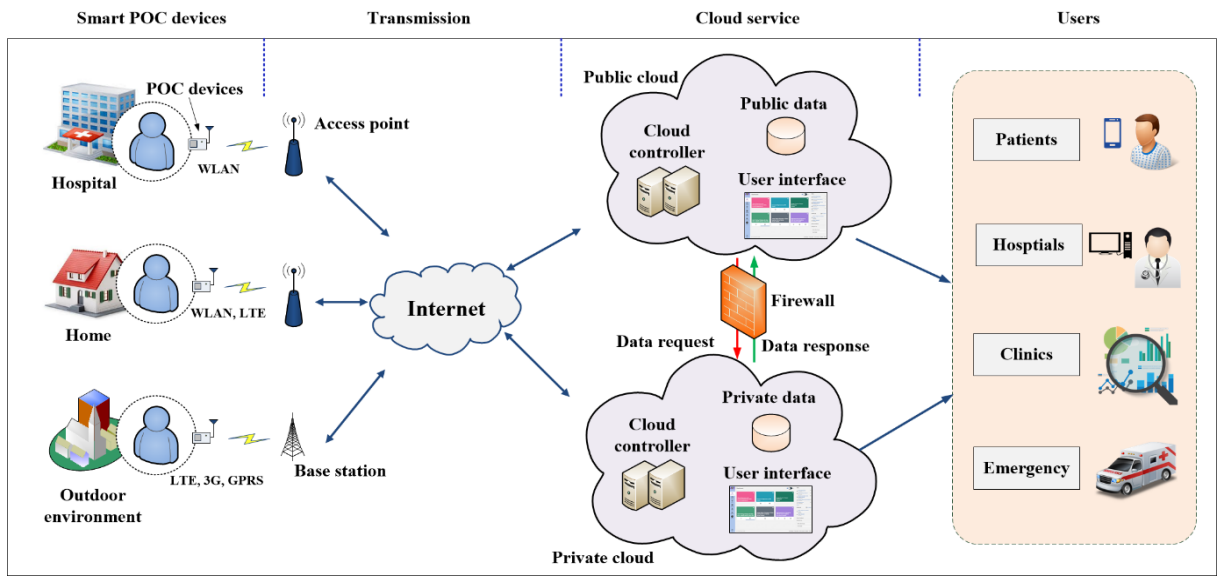

**Fig. 14** A framework of pervasive health diagnosis system [181]

Such a pervasive diagnosis system based on the POC devices and the IoT platform does not involve the use of laboratory staff and facilities, and can provide access to health monitoring and assessment technologies for people with limited or no healthcare facilities, or with geographically distance or difficulty to physically access facilities. By now, the classes of analytical targets in spintronic biodetection are expanding from proteins and nucleic acids to metabolites, drugs, and even cells [182]. By testing the samples which are easy to get, such as blood, saliva, and urine, the POC testing can get various health information "near-patient" whenever it is needed. Meanwhile, the POC diagnostics are widely self-administered without the need of user' complex operations, and thus patients can conveniently realize their health conditions at home or outdoor environment, reducing the frequency of hospital visits and travel expenses. This kind of spintronic POC system will be robust, low-cost, with no requirement for maintenance or calibration, and use wireless to transmit data. All these are favorable for its pervasiveness in smart healthcare.

Hospitals and healthcare facilities can provide high-efficient healthcare services benefiting from this pervasive diagnosis system. The monitored information of patients can be remotely accessed by the designated hospitals. Such information could be useful for maintaining patients' health, managing disease, or monitoring therapy. Hence, doctors or healthcare staff can directly provide professional help to patients over the Internet. Meanwhile, user's profile and medical history data are maintained by the management center for local private use. The doctor may access the user's information as needed when making the clinical decisions. Moreover, automated notifications or alarms can be issued to his/her relatives according to the diagnosis results via various telecommunication means. In addition, EMS also benefit from this pervasive diagnosis system. Traditional EMS starts with a 911 call followed by ground ambulance dispatch, patient evaluation, treatment by EMS personnel, and transport to a hospital facility [183]. Time available for EMS patient evaluation and treatment varies on the basis of geography and patient condition. The pervasive diagnosis system can offer the long-term and recent diagnosis records of the patient. This will be beneficial to the clinical decision-making and EMS arrangement at early stage, shortening the waiting time.

This cloud-enabled pervasive healthcare system shares the long-term monitored data with authorized social network or medical research communities, which can search for personalized medical trends or group patterns. In this way, the life-threatening events could be detected and controlled in the early stage. Furthermore, the cloud services can also provide insights into the disease evolution, the rehabilitation process, and the effect of drug therapy. The longitudinal monitoring of key biomarkers of a certain disease (e.g. cancer) from all users is expected to provide representative information about related environmental factors, distribution susceptible population, change of incidence, and prognosis of different therapy strategies. It is reasonable to foresee that, by being integrated into a smart lifestyle, this pervasive diagnosis system will greatly improve the quality of citizens' healthcare.

## V. Conclusion and outlook

This review focuses on the promising combination of spintronic sensors and IoT that enables the realization of smart living. Spintronic sensors, working as sensing devices, possess superb measuring abilities, and they are of small size and low cost. They can be used as pervasive magnetic field sensors in electrical current sensing, transmission and distribution lines monitoring, vehicle detection and biodetection. By virtue of the IoT technologies, spintronic sensors are seamlessly integrated into WSNs, which provide the pervasive WSSN monitoring systems for smart living. These WSSN based or related solutions enable the pervasive building energy management system, the wide-area transmission and distribution network monitoring system, the all-round traffic monitoring and management system, and the pervasive health diagnosis system. Indeed, the applications of WSSNs are not limited to the domains above. In the future, the development of spintronic sensors with wider linear measuring range, higher sensitivity, and lower noise will gain more opportunities in sensing ultra-weak or transient signals (e.g., in detecting low-concentration target analyte molecules and tracking high-speed motion). Note that there is no single universal sensor solution to all sensing problems. Nevertheless, the unique advantages and features of spintronic sensors, combining together with other competing or complementary sensing technologies, will go a long way in stimulating the advancement of smart living in all aspects. Multiple-source data fusion will combine spintronic sensors with other sensing technologies such as environmental sensors to reach a comprehensive platform to extend smart sensing into even broader regime. Meanwhile, the advancement of cloud computing and big data technologies will hold great promise to provide trend prediction based on the historical data, facilitating the policy making in society or infrastructure at the early stage.

**Acknowledgements**

This work was supported in part by the Seed Funding Program for Basic Research, Seed Funding Program for Applied Research and Small Project Funding Program from the University of Hong Kong, ITF Tier 3 funding (ITS/214/14), and University Grants Committee of Hong Kong (Contract No. AoE/P-04/08).